\pdfoutput=1
\documentclass{article}
\usepackage{arxiv} 
\usepackage[utf8]{inputenc} % allow utf-8 input
\usepackage[T1]{fontenc}    % use 8-bit T1 fonts
\usepackage{hyperref}       % hyperlinks
\usepackage{url}            % simple URL typesetting
\usepackage{booktabs}       % professional-quality tables
\usepackage{amsfonts}       % blackboard math symbols
\usepackage{nicefrac}       % compact symbols for 1/2, etc.
\usepackage{microtype}      % microtypography
\usepackage{lipsum}
\usepackage{graphicx}
\usepackage{booktabs}  
\usepackage{multirow} 
\usepackage{array}

\usepackage{csquotes}
\usepackage{authblk}
\usepackage[labelsep=period]{caption}

\graphicspath{ {./images/} }
\title{{EVA-MED: An Enhanced Valence-Arousal Multimodal Emotion Dataset for Emotion Recognition}}

\author{
    Xin Huang$^{1,2*}$\thanks{Corresponding author: hx11@pku.edu.cn}, Shiyao Zhu$^{1,2}$, Ziyu Wang$^{1,2}$, Yaping He$^{3,4}$, Hao Jin$^{1,2*}$, Zhengkui Liu$^{3,4*}$ \\  
    \small
    $^1$ College of Information Science and Electronic Engineering, Zhejiang University, Hangzhou, China \\
    $^2$ International Joint Innovation Center, Zhejiang University, Haining, China \\
    $^3$ CAS Key Laboratory of Mental Health, Institute of Psychology \\
    $^4$ Department of Psychology, University of Chinese Academy of Sciences \\
    * Corresponding author: Xin Huang, E-mail: hx11@pku.edu.cn
}

\begin{document}
\maketitle
\begin{abstract}
We introduce a novel multimodal emotion recognition dataset that enhances the precision of Valence-Arousal Model while accounting for individual differences.
This dataset includes electroencephalography (EEG), electrocardiography (ECG), and pulse interval (PI) from 64 participants. 
Data collection employed two emotion induction paradigms: video stimuli that targeted different valence levels (positive, neutral, and negative) and the Mannheim Multicomponent Stress Test (MMST), which induced high arousal through cognitive, emotional, and social stressors.
To enrich the dataset, participants' personality traits, anxiety, depression, and emotional states were assessed using validated questionnaires. 
By capturing a broad spectrum of affective responses while accounting for individual differences, this dataset provides a robust resource for precise emotion modeling.
The integration of multimodal physiological data with psychological assessments lays a strong foundation for personalized emotion recognition. 
We anticipate this resource will support the development of more accurate, adaptive, and individualized emotion recognition systems across diverse applications.
\end{abstract}

\section{Background and Summary}

Emotion recognition plays a vital role in various fields, including mental health support, human-computer interaction, education, and marketing. 
By accurately identifying and measuring emotional states, emotion recognition has the potential to create more personalized experiences, enhance user engagement, and support mental health and well-being \cite{cai_emotion_2023}.

For instance, in mental health, emotion recognition could monitor fluctuations in an individual's mood in real-time, enabling the early detection of risks for psychological disorders such as depression and anxiety \cite{cohen_psychiatric_2013}, which allows for timely interventions.
In recent years, advancements in the interdisciplinary fields of affective computing and neuroscience have significantly accelerated the development of emotion recognition technology \cite{saganowski_emognition_2022}. Furthermore, progress in deep learning algorithms and multimodal data fusion has greatly improved the accuracy and adaptability of emotion recognition systems \cite{ezzameli_emotion_2023}.

Existing multimodal emotion datasets, such as DEAP \cite{koelstra_deap_2012}, AMIGOS \cite{miranda_calero_wemac_2024}, and SEED-VII \cite{jiang_seed-vii_2024}, have significantly advanced the field by integrating diverse physiological signals, including electroencephalography (EEG), electrocardiography (ECG), and self-reported emotional labels. 
These datasets have enabled the development of machine learning models that can recognize complex emotional states. However, several limitations still exist \cite{ramaswamy_multimodal_2024}. 
One major issue is that most current emotion studies rely on the continuous emotional model to differentiate between valence but do not address the distinctions in arousal \cite{katsigiannis_dreamer_2018, miranda_calero_wemac_2024}. 
Additionally, while individual differences such as personality traits and anxiety levels are known to significantly impact emotional processing, there is a lack of comprehensive datasets that systematically incorporate these factors \cite{ramaswamy_multimodal_2024}. 
This gap restricts the depth of analysis, particularly in understanding how personal characteristics influence emotional responses under various conditions.

To address these limitations, we present a novel multimodal emotion recognition dataset designed to enhance the precision of emotional dimension modeling and systematically account for individual differences. 
Our approach is grounded in the widely accepted two-dimensional model of emotion, which conceptualizes emotions along the orthogonal dimensions of valence and arousal \cite{zhao_emotion_2021}. 
To comprehensively capture emotional variability, we employed two complementary emotion elicitation paradigms: video-based emotion induction and the Mannheim Multicomponent Stress Test (MMST). 
While both paradigms collect data on valence and arousal, they exhibit distinct strengths in eliciting specific emotional responses. 
Video-based tasks are particularly effective in inducing a range of emotional valence, including positive, negative, and neutral states, while simultaneously triggering moderate levels of arousal. 
In contrast, the MMST is designed to evoke high arousal levels through a combination of stress-inducing components, such as time pressure, negative feedback, and cognitive load, while also eliciting emotional valence shifts associated with stress responses \cite{reinhardt_salivary_2012}. 
This dual-paradigm approach ensures broad coverage of emotional states, ranging from low to high arousal and spanning the full spectrum of valence (positive, neutral, and negative), thereby enhancing ecological validity and dataset diversity.

In addition to comprehensive emotional dimension modeling, our dataset integrates assessments of individual differences to explore their influence on emotional processing. 
Participants completed a series of psychometrically validated questionnaires designed to measure personality traits, anxiety, depression, and life events. 
By incorporating these individual characteristics, the study enables a deeper analysis of how physiological and behavioral responses are modulated by emotional stimuli, providing a richer context for emotion recognition research.

Regarding data collection, we combined high-precision physiological recording techniques with exploratory applications of wearable device technology. 
EEG and ECG data were collected using laboratory-grade equipment, whereas pulse interval (PI) data were obtained from wrist-worn wearable devices.
Although wearable technology is currently constrained by accuracy and data resolution limitations, we acknowledge its potential for future large-scale data collection.
The portability and ease of use of wearable devices create possibilities for emotion data acquisition in real-world settings, enabling long-term, low-intrusion monitoring of emotional dynamics \cite{wijasena_survey_2021,lee_current_2021}. 
This capability not only enhances the ecological validity of emotion recognition models but also supports the development of personalized emotion-aware technologies, such as emotion-adaptive interfaces and mental health monitoring tools \cite{kwon_emotion_2021,shu_wearable_2020}.

In summary, this novel multimodal emotion recognition dataset is distinguished by its precise modeling of emotional dimensions and its comprehensive consideration of individual differences. 
The integration of controlled laboratory data with exploratory wearable device applications lays the foundation for future scalability and real-world applicability. 
We believe this dataset will significantly contribute to the advancement of emotion recognition research, fostering the development of more accurate, personalized, and context-aware affective computing systems.

\section{Methods}
\label{sec:headings}
\subsection{Participants}
A total of 64 university students (33 males and 31 females) participated in this study, with ages ranging from 19 to 26 years (M = 22.06, SD = 2.08). 
Participants were recruited through advertisements at universities in Beijing. 
All individuals were physically and mentally healthy. 
Inclusion criteria required participants to be over 18 years old, enrolled in college, not currently using psychotropic medications, and free from significant neurological or cardiovascular conditions. 
Exclusion criteria included the use of psychiatric medications within the past six months, a diagnosis of major mental health disorders such as schizophrenia or major depressive disorder, and any physiological abnormalities affecting cardiac function. 
The study was approved by the Ethics Committee of the Institute of Psychology, Chinese Academy of Sciences, and all participants provided informed consent before their involvement.

\subsection{Procedure}
The experiment was divided into three distinct phases (see Figure 1). 
In the initial phase, participants were instructed to relax while completing a questionnaire and undergoing physiological data collection. 
This phase, corresponding to a “calm” state, served as the baseline reference for subsequent emotional change.

In the second phase, participants underwent the emotion induction phase, which involved watching video clips corresponding to different emotional valences (positive, neutral, negative). 
For the first thirty participants, the viewing order was positive, neutral, and then negative; for the remaining participants, the order was reversed to negative, neutral, and then positive. 
This counterbalancing minimized potential carryover effects between emotional states. 
At the end of each video, participants completed Self-Assessment Manikin (SAM) ratings to evaluate both valence and arousal, providing subjective reports of their emotional responses to the viewed materials. 

In the third phase, participants first watched a two-minute neutral video to stabilize their emotional state following the emotion induction phase. 
They then proceeded to perform the MMST, during which SAM ratings were obtained at 0, 3, 5, and 8 minutes, while EEG, ECG, and PI data were continuously recorded.

\begin{figure}[h] 
  \centering  
  \includegraphics[width=0.9\textwidth]{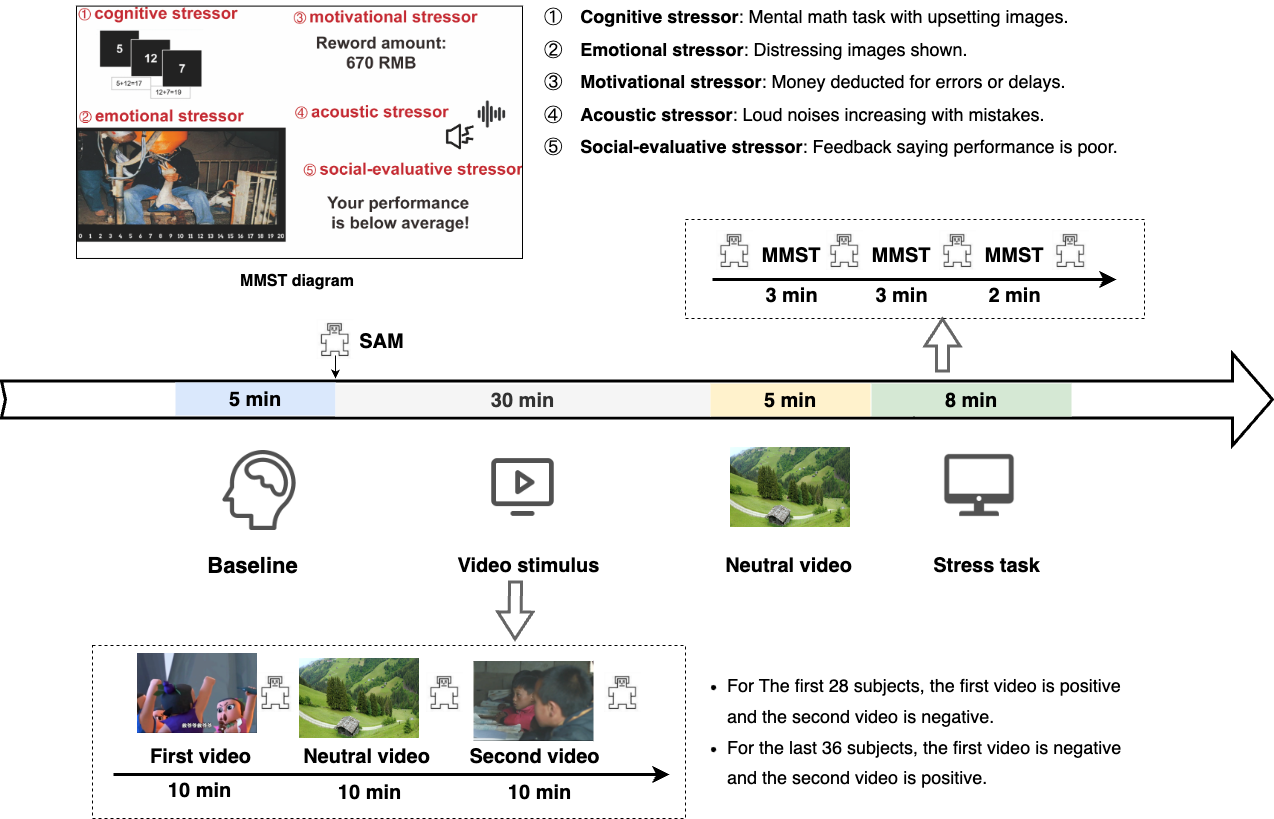}
  \caption{Experiment Procedures.} 
  \label{fig1} 
\end{figure}

\subsection{Stimuli}
\subsubsection{Video clips}
To ensure emotional efficacy, a pre-experiment was conducted with 55 college students who rated the emotional intensity and differentiation of candidate clips. 
The final selections demonstrated strong emotional induction effects, validated through subjective ratings on a four-point Likert scale.

The video materials were classified into three categories according to emotional valence, with each video lasting for 10 minutes.

\begin{itemize}
  \item Positive: humorous short clips sourced from Chinese internet media platforms to evoke positive emotions.
  \item Neutral: videos depicting everyday objects and landscapes with neutral background music to maintain a baseline emotional state.
  \item Negative: documentary footage featuring interviews with left-behind children to elicit negative emotions.
\end{itemize}

\subsubsection{MMST paradigm}
The Mannheim Multicomponent Stress Test (MMST) combines five distinct stressors to elicit a heightened state of arousal. 
Cognitive stress was induced using the Paced Auditory Serial Addition Test (PASAT) \cite{lejuez_modified_2003}, in which participants added consecutive numbers and selected the correct answer from 21 options within approximately 3.5 seconds per trial. 
Real-time feedback was provided throughout the 8-minute task. 
Additionally, participants performed the arithmetic task while emotionally aversive images from the Geneva Affective Picture Database (GAPED) \cite{dan-glauser_geneva_2011} were presented, thereby incorporating an emotional stressor. 
Acoustic stress was introduced via countdown signals and explosion-like noises that intensified after incorrect responses. A motivational stressor was implemented by deducting monetary amounts from an initial balance for incorrect or delayed responses. 
Finally, social evaluative stress was applied through on-screen feedback, especially at the third and fifth minutes, indicating that performance was below average compared to peers. To further increase task difficulty, only responses made within a designated central area on the screen were accepted; 
any responses outside this area were deemed invalid regardless of correctness.

\subsection{Measures}
Participants completed a series of questionnaires and self-assessment scales to evaluate demographic information, personality traits, anxiety, depression.

\subsubsection{Demographic Information Questionnaire}
Collected basic demographic data, including gender, age, education level, family socioeconomic status, physical health status, and relationship status.

\subsubsection{Psychosocial Measures}
\begin{itemize}
  \item Chinese Big Five Personality Inventory-15 (CBF-PI-15): assessed personality traits across the five dimensions, openness, conscientiousness, extraversion, agreeableness, and neuroticism, using 15 items rated on a 6-point scale (1 = strongly disagree, 6 = strongly agree) \cite{zhang_development_2019}.
  \item Patient Health Questionnaire-9 (PHQ-9): measured the severity of depressive symptoms through 9 items rated on a 4-point scale (0 = not at all, 3 = nearly every day) \cite{wang_reliability_2014}.
  \item State-Trait Anxiety Inventory (STAI): evaluated trait anxiety levels through 20 items rated on a 4-point scale (1 = seldom, 4 = almost always) \cite{spielberger_state-trait_2017}.
  \item Adolescent Self-Rating Life Events Checklist (ASLEC): assessed the frequency and intensity of recent life events and associated stress with 27 items rated on a 6-point scale (0 = not at all, 5 = extremely) \cite{liu_adolescent_1997}.
\end{itemize}

\subsubsection{Subjective Emotional Experience}
SAM: quantified participants' current emotional valence and arousal levels \cite{bradley_measuring_1994}, using a graphical representation with a 9-point scale, later converted to a 1-5 scale in increments of 0.5, shown in Figure 2.

\begin{figure}[h] 
  \centering  
  \includegraphics[width=0.6\textwidth]{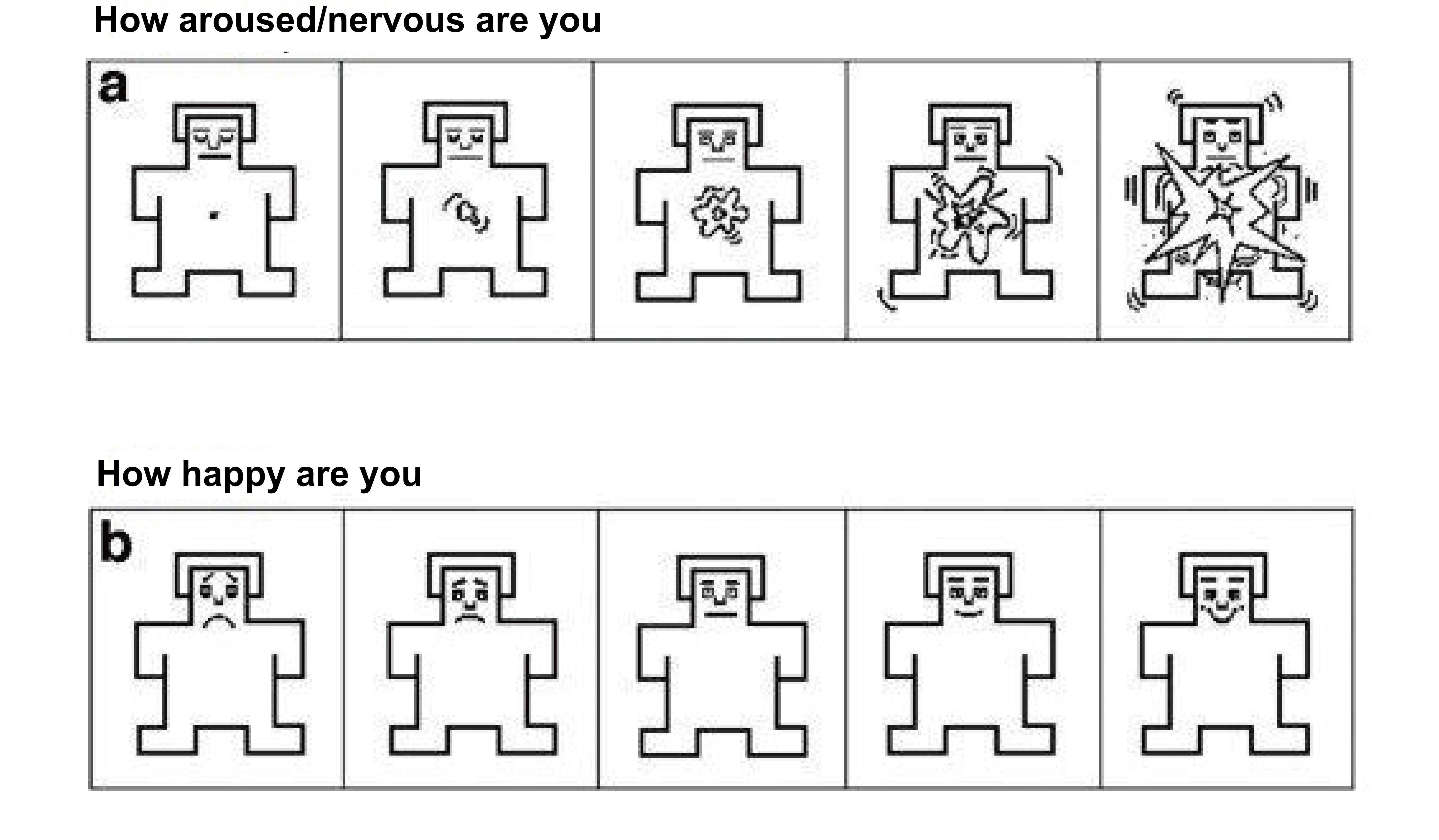}
  \caption{SAM Diagram.} 
  \label{fig2} 
\end{figure}

\subsection{Equipment}

\begin{figure}[h]
  \centering
  \includegraphics[width=0.6\textwidth]{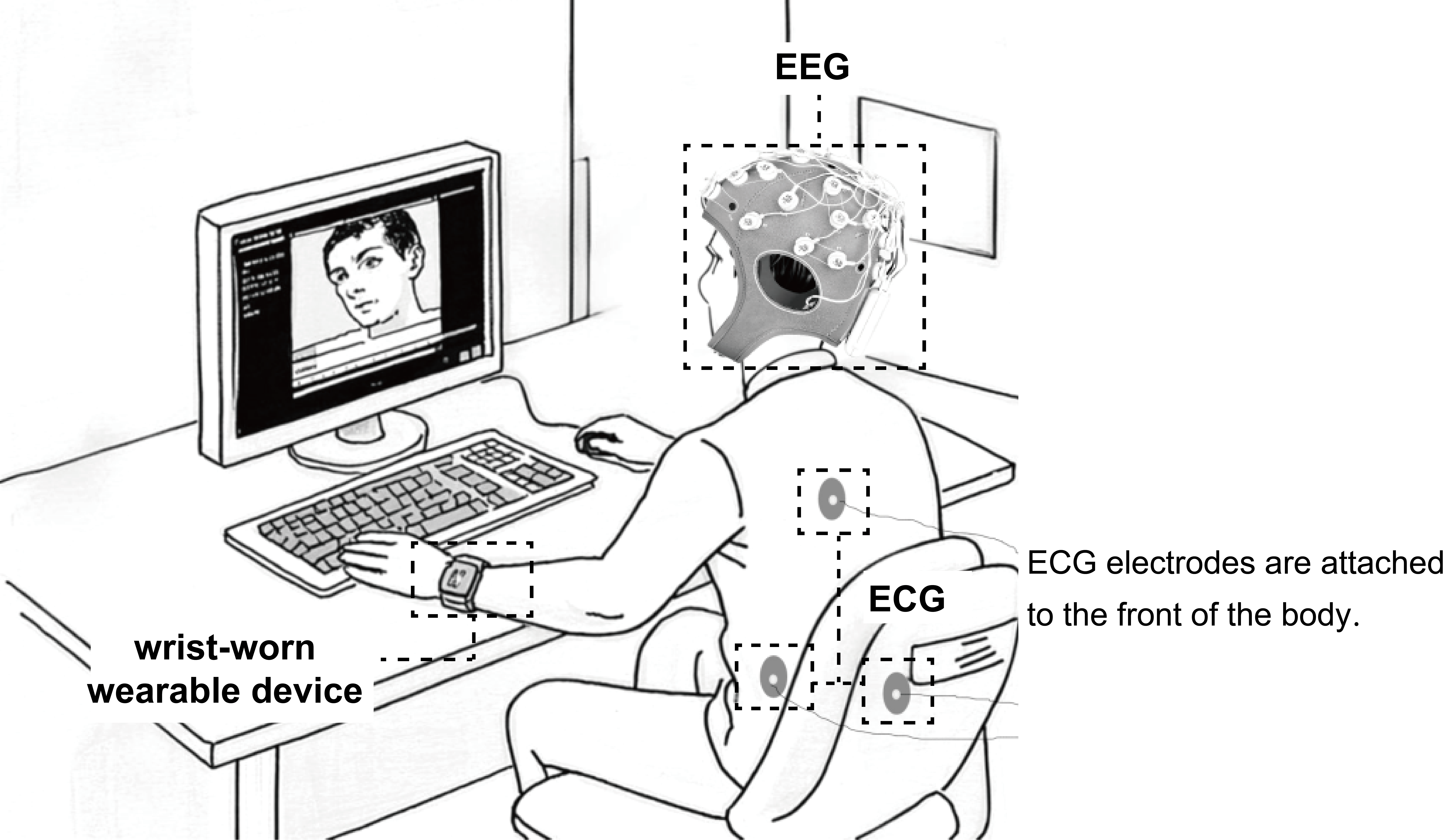}
  \caption{Experimental Setup for Multimodal Physiological Data Collection. Participants wore an EEG cap to record brain activity, while ECG electrodes were placed on the chest and lower rib area to monitor heart activity. A wrist-worn wearable device was used to collect pulse interval (PI) data. The participant performed tasks on a computer while physiological signals were continuously recorded.} 
  \label{fig3}
\end{figure}

\subsubsection{EEG Equipment}
EEG signals were collected using the NeuroScan system with 64 electrodes arranged according to the international 10-20 electrode placement system. 
The online reference electrode was placed on the left mastoid, while the offline reference was the arithmetic average of the bilateral mastoids. 
The ground electrode was positioned between FPz and Fz. Vertical eye movement electrodes were placed above and below the left eye, and horizontal eye movement electrodes were placed laterally on both sides of the eyes. 
Data collection and real-time monitoring were performed using Scan 4.5 software.

\subsubsection{ECG Equipment}
ECG data were collected using the Biopac MP150 system equipped with an ECG module, which recorded single-channel ECG signals. 
Before data acquisition, the skin was cleaned with alcohol and the disposable electrodes were affixed to three locations: the right clavicular midline at the sternal notch level, the left lower rib middle area, and the right lower abdominal area (ground). 
Data were captured using AcqKnowledge 4.2 software at a sampling rate of 2000 Hz.

\subsubsection{PI Data Collection}

PI refers to the time between consecutive pulse waves, reflecting the variability in heartbeats over time (See Figure 4). 
PI data were collected using the consumer-grade Huawei Band 7. The device sampled data at 25 Hz, capturing data on a per-minute basis. The data were transmitted via Bluetooth for further analysis.

\begin{figure}[h] 
  \centering  
  \includegraphics[width=0.35\textwidth]{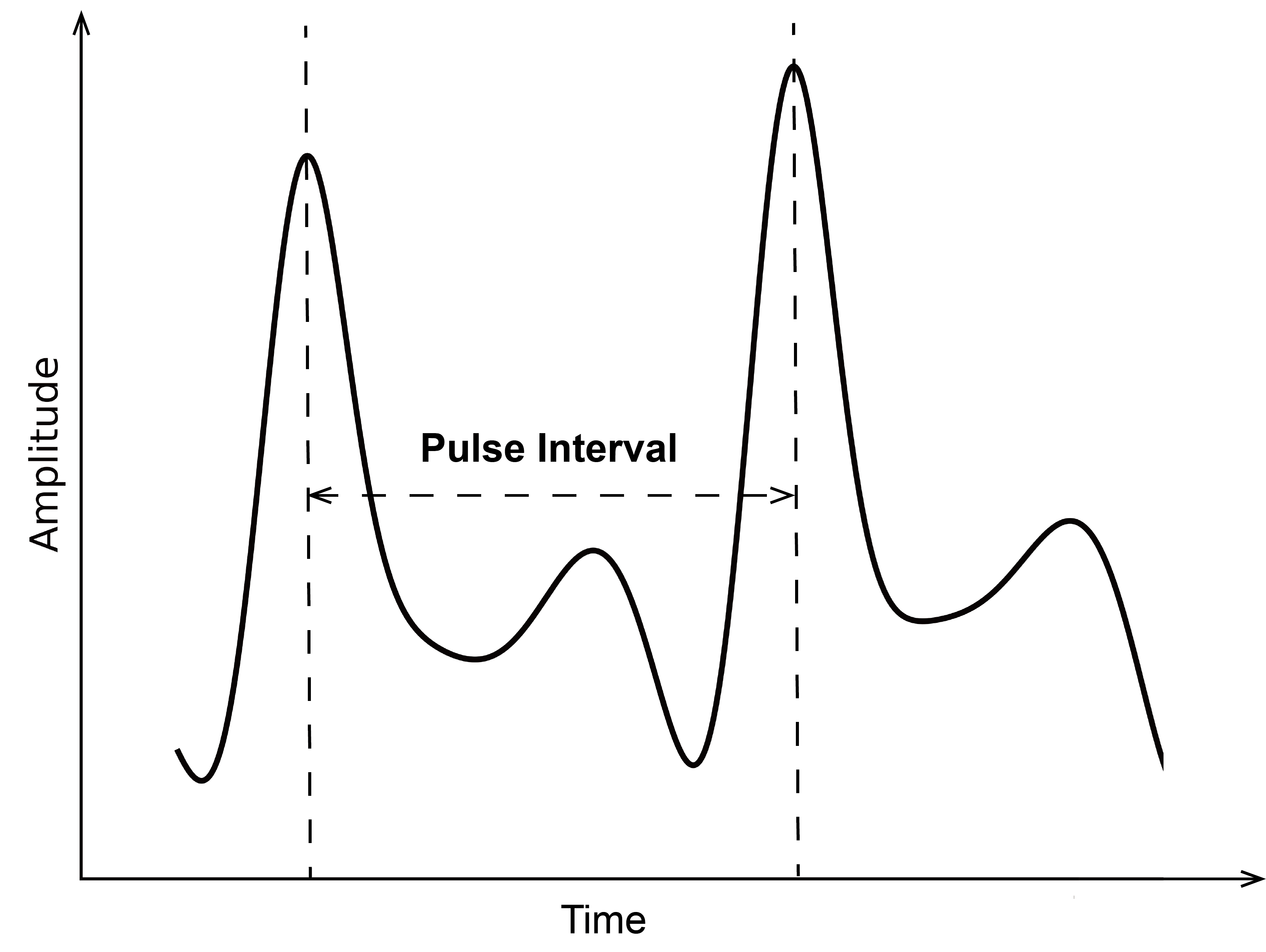}
  \caption{Pulse Interval.} 
  \label{fig4}  
\end{figure}

\subsubsection{Software for Data Processing}
Data preprocessing was conducted using Python with the following packages. 
EEG and ECG data preprocessing were performed using the MNE toolkit, R-R interval (RRI) values were extracted from the ECG data using NumPy, outlier filtering for PI data was done using the HRV package, and the analysis of PI values was carried out using the NeuroKit2 package (Makowski et al., 2021).

\subsection{Data Preprocessing}
The dataset comprises EEG, ECG, and PI signals collected from 64 participants. 
All data segments are meticulously labeled with valence and arousal scores to facilitate comprehensive emotion recognition tasks. 
The preprocessing pipeline ensures data quality, consistency, and precise temporal alignment across the different physiological modalities.

\subsubsection{EEG Data Preprocessing}
Initially, spherical spline interpolation was performed to correct any bad channels in the EEG recordings. 
Subsequently, the data were filtered using a fourth-order zero-phase Butterworth bandpass filter with a frequency range of 0.5 Hz to 45 Hz to remove unwanted noise. 
Artifact-contaminated segments were then automatically identified and removed based on excessive high-frequency activity. Next, Independent Component Analysis (ICA) was applied to further eliminate artifacts such as eye blinks, cardiac signals, and movement-related noise. 
Finally, the cleaned EEG data were segmented into 4-second epochs according to the emotional labels and synchronized with the ECG data.

\subsubsection{ECG Data Preprocessing}
Initially, the raw ECG data were filtered using a high-pass filter at 0.5 Hz, a 5th-order Butterworth filter, and a notch filter at 50 Hz to eliminate power line interference. 
Then the data were segmented based on emotional labels into 4-second intervals for subsequent statistical analysis. 
R-wave detection was performed based on the absolute gradient of the ECG signals, identifying the peaks corresponding to heartbeats. 
Subsequently, RRI, defined as the time between consecutive R-wave peaks, was calculated for further analysis.

\subsubsection{PI Data Preprocessing}
Anomaly filtering was performed to eliminate outliers and artifacts. Due to the minute-by-minute sampling rate of the PI data, only short-term heartbeat dynamic features could be extracted, limiting the analysis of long-term heart rate variability metrics.

\section{Data Records}
\label{sec:others}
The dataset was organized into four sections, each with a README.txt file explaining its contents.

\textbf{Physiological Data}: This folder contains three subfolders.
\begin{itemize}
  \item Collected Data: This folder contains three subfolders, each dedicated to a specific type of data:   raw EEG recordings in .cnt format, multi-channel physiological data in .acq format, and raw PI data collected via a wearable device.
  \item Labeled Data: This folder contains two subfolders, one folder contains EEG and ECG data segmented by emotional states and time-aligned, stored in .npy format, while another contains PI data segmented by emotional states for each subject.
  \item Preprocessed Data: this folder contains three subfolders, one storing preprocessed ECG data in .npy format, another storing preprocessed EEG data in .fif format. Both datasets are segmented into 4-second intervals based on emotional labels and time-synchronized. The third subfolder contains preprocessed PI data.
\end{itemize}

\textbf{Questionnaire Data}: This folder contains two .xlsx file.
\begin{itemize}
  \item Demographic\_Information.xlsx: This file contains demographic details of all participants, with each row corresponding to an individual subject.
  \item SAM.xlsx: This file stores SAM ratings, documenting valence and arousal scores across different experimental phases.
\end{itemize}

\textbf{Channel Order}: This folder contains two .xlsx file.
\begin{itemize}
  \item EEG\_order.xlsx: This file provides the EEG channel order information.
  \item EEG\_ECG\_order.xlsx: This file provides the time-aligned EEG and ECG mentioned in Labeled Data. 
\end{itemize}

\textbf{Stimuli}: This folder contains three .mp4 file.
\begin{itemize}
  \item Positive\_video.mp4: short humorous clips sourced from Chinese media platforms.
  \item Neutral\_video.mp4: scenic footage with soft instrumental music.
  \item Negative\_video.mp4: documentary excerpts featuring interviews with left-behind children.
\end{itemize}

\begin{figure}[h] 
  \centering  
  \includegraphics[width=0.9\textwidth]{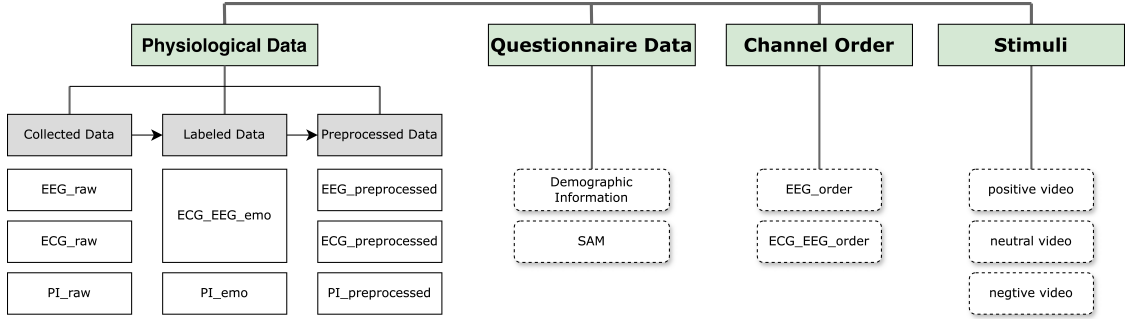}
  \caption{Dataset Structure.} 
  \label{fig5} 
\end{figure}

\section{Technical Validation}
In this study, an emotion recognition prediction model was established for the two dimensions: emotional valence and emotional arousal. 
Emotional valence ratings on the SAM scale were categorized into three classes: negative (1-2 points), neutral (3 points), and positive (4-5 points). 
In addition, an independent label calm was considered, resulting in four overall valence labels. 
Emotional arousal was transformed into calm (1-2 points) and stress (4-5 points) as the two emotional arousal labels. Multiple machine learning and deep learning algorithms were used for modeling.

Initially, features were extracted from the preprocessed data. For EEG signals, Principal Component Analysis (PCA) was applied to reduce spatial dimensionality, and the first 10 principal components were retained. 
The cumulative variance explained by the principal components reached 75\%.
Frequency-domain features as well as time-domain and time-frequency features, were computed for each component, yielding a total of 80 features. 
RRI was extracted from ECG data, and a comprehensive set of 94 features was computed using time-domain, frequency-domain, and nonlinear analysis.
An identical feature extraction procedure was applied to the PI data; however, due to the per-minute aggregation of PI data, the analysis was restricted to short-term heart rate dynamics, resulting in 78 features.

Next, predictive models were constructed for emotional valence and emotional arousal using multiple machine learning and deep learning algorithms. 
The models included traditional machine learning techniques such as Support Vector Machines (SVM), Random Forest, Decision Trees, and Extreme Gradient Boosting (XGBoost), along with deep learning models such as Multi-Layer Perceptron (MLP) and 1D Convolutional Neural Networks (1dCNN).
The dataset was split into 80\% for training and 20\% for testing, with 10-fold cross-validation used to assess model generalizability. 
Model performance was evaluated using accuracy and F1-score.

\begin{table}[ht]
    \centering
    \renewcommand{\arraystretch}{1.2} 
    \begin{tabular}{l l c c c c c c}
        \toprule
        \multirow{2}{*}{Data Type} & \multirow{2}{*}{} & 
        \multicolumn{2}{c}{EEG} & \multicolumn{2}{c}{ECG} & \multicolumn{2}{c}{PI} \\
        \cmidrule(lr){3-4} \cmidrule(lr){5-6} \cmidrule(lr){7-8}
        & & ACC & F1-score & ACC & F1-score & ACC & F1-score \\
        \midrule
        \multirow{2}{*}{SVM} & Valence & 74.00\% & 0.73 & 45.08\% & 0.45 & 48.78\% & 0.48 \\
                             & Arousal & 77.12\% & 0.75 & 70.83\% & 0.70 & 64.86\% & 0.73 \\
       
        \multirow{2}{*}{Decision Tree} & Valence & 61.89\% & 0.61 & 45.87\% & 0.46 & 37.01\% & 0.36 \\
                                       & Arousal & 72.74\% & 0.69 & 76.67\% & 0.76 & 57.84\% & 0.69 \\
       
        \multirow{2}{*}{Random Forest} & Valence & 81.72\% & 0.81 & 33.90\% & 0.33 & 48.47\% & 0.47 \\
                                       & Arousal & 80.97\% & 0.76 & 75.83\% & 0.75 & 65.96\% & 0.77 \\
       
        \multirow{2}{*}{XGBoost} & Valence & 80.69\% & 0.80 & 42.11\% & 0.41 & 50.12\% & 0.49 \\
                                 & Arousal & 90.17\% & 0.89 & 81.67\% & 0.81 & 67.00\% & 0.76 \\
        
        \multirow{2}{*}{MLP} & Valence & 44.51\% & 0.43 & 67.03\% & 0.56 & 43.12\% & 0.44 \\
                             & Arousal & 77.10\% & 0.67 & 82.50\% & 0.82 & 63.17\% & 0.74 \\
        
        \multirow{2}{*}{1dCNN} & Valence & 90.46\% & 0.90 & 81.60\% & 0.82 & 73.23\% & 0.76 \\
                               & Arousal & 93.44\% & 0.91 & 86.65\% & 0.89 & 73.92\% & 0.76 \\
        \bottomrule
    \end{tabular}
    \caption{Classification Performance of Emotion Valence and Arousal Across Different Modalities (EEG, ECG, and PI).}
    \label{tab:classification_performance}
\end{table}

\begin{figure}[h] 
  \centering  
  \includegraphics[width=0.85\textwidth]{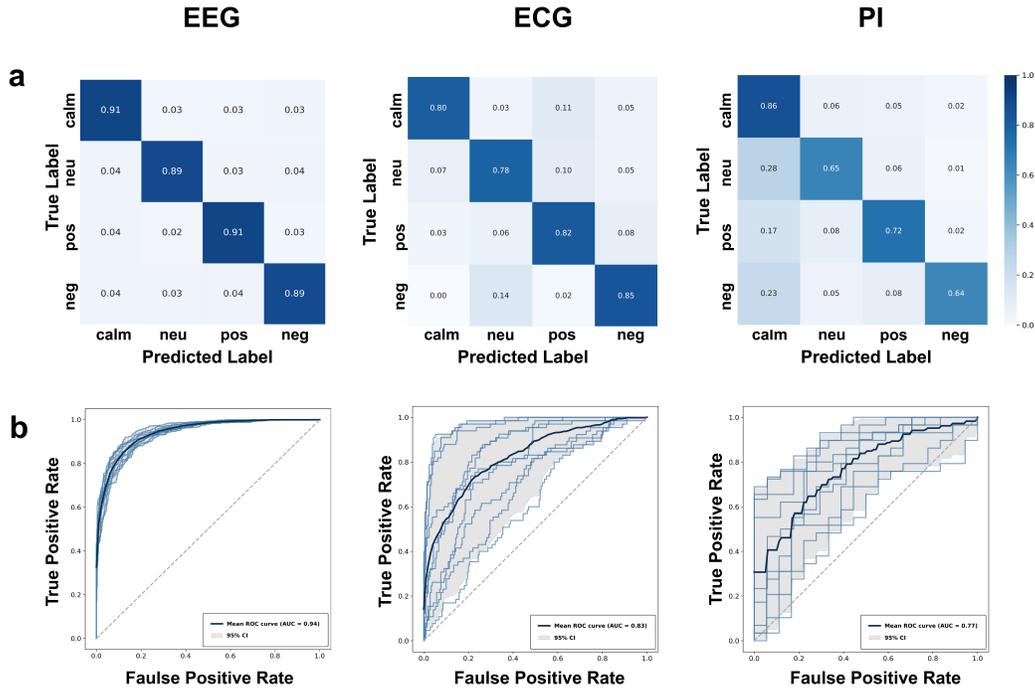}
  \caption{1dCNN Models for Emotional Valence and Arousal. (a) Confusion matrix for emotional valence prediction by EEG, ECG, and PI features. (b) ROC curve for emotional arousal prediction by EEG, ECG, and PI features.} 
  \label{fig6} 
\end{figure}

Notably, the 1dCNN model outperformed traditional machine learning models in predicting both emotional valence and arousal, with accuracies of 90.46\% and 93.44\% for the EEG-based model, 
82.00\% and 86.65\% for the ECG-based model, and 73.23\% and 73.92\% for the PI-based model, respectively. 
Despite the moderate performance of the PI-based models, their suitability for long-term, low-intrusion monitoring underscores their practical applicability in real-world emotion recognition systems.

In this study, we present a comprehensive multimodal emotion recognition dataset aimed at improving the precision of Valence-Arousal Model while systematically incorporating individual differences. By integrating EEG, ECG, and PI signals collected under controlled emotional stimuli, this dataset serves as a valuable resource for advancing research in emotion recognition and computational affective science.

Future studies can expand the dataset to include a larger and more diverse population across different age groups, cultural backgrounds, and real-world scenarios. Furthermore, enhancing the real-time performance and robustness of emotion recognition systems is crucial. By refining data diversity and enhancing real-time processing capabilities, emotion recognition systems can achieve enhanced accuracy and applicability in everyday settings and human-computer interaction.

\bibliographystyle{plain}  % 选择一种引用样式，这里使用 plain 样式
\bibliography{references}  % 引用的 .bib 文件，确保在提交时包含 references.bib 文件

\begin{thebibliography}{10}

\bibitem{bradley_measuring_1994}
Margaret~M. Bradley and Peter~J. Lang.
\newblock Measuring emotion: {The} self-assessment manikin and the semantic
  differential.
\newblock {\em Journal of Behavior Therapy and Experimental Psychiatry},
  25(1):49--59, March 1994.

\bibitem{cai_emotion_2023}
Yujian Cai, Xingguang Li, and Jinsong Li.
\newblock Emotion {Recognition} {Using} {Different} {Sensors}, {Emotion}
  {Models}, {Methods} and {Datasets}: {A} {Comprehensive} {Review}.
\newblock {\em Sensors}, 23(5):2455, January 2023.

\bibitem{cohen_psychiatric_2013}
Alex~S. Cohen, Yunjung Kim, and Gina~M. Najolia.
\newblock Psychiatric symptom versus neurocognitive correlates of diminished
  expressivity in schizophrenia and mood disorders.
\newblock {\em Schizophrenia Research}, 146(0):249--253, May 2013.

\bibitem{dan-glauser_geneva_2011}
Elise~S. Dan-Glauser and Klaus~R. Scherer.
\newblock The {Geneva} affective picture database ({GAPED}): a new 730-picture
  database focusing on valence and normative significance.
\newblock {\em Behavior Research Methods}, 43(2):468--477, June 2011.
\newblock Company: Springer Distributor: Springer Institution: Springer Label:
  Springer Number: 2 Publisher: Springer-Verlag.

\bibitem{ezzameli_emotion_2023}
K.~Ezzameli and H.~Mahersia.
\newblock Emotion recognition from unimodal to multimodal analysis: {A} review.
\newblock {\em Information Fusion}, 99:101847, November 2023.

\bibitem{jiang_seed-vii_2024}
Wei-Bang Jiang, Xuan-Hao Liu, Wei-Long Zheng, and Bao-Liang Lu.
\newblock {SEED}-{VII}: {A} {Multimodal} {Dataset} of {Six} {Basic} {Emotions}
  with {Continuous} {Labels} for {Emotion} {Recognition}.
\newblock {\em IEEE Transactions on Affective Computing}, pages 1--16, 2024.
\newblock Conference Name: IEEE Transactions on Affective Computing.

\bibitem{katsigiannis_dreamer_2018}
Stamos Katsigiannis and Naeem Ramzan.
\newblock {DREAMER}: {A} {Database} for {Emotion} {Recognition} {Through} {EEG}
  and {ECG} {Signals} {From} {Wireless} {Low}-cost {Off}-the-{Shelf} {Devices}.
\newblock {\em IEEE Journal of Biomedical and Health Informatics},
  22(1):98--107, January 2018.
\newblock Conference Name: IEEE Journal of Biomedical and Health Informatics.

\bibitem{koelstra_deap_2012}
Sander Koelstra, Christian Muhl, Mohammad Soleymani, Jong-Seok Lee, Ashkan
  Yazdani, Touradj Ebrahimi, Thierry Pun, Anton Nijholt, and Ioannis Patras.
\newblock {DEAP}: {A} {Database} for {Emotion} {Analysis} ;{Using}
  {Physiological} {Signals}.
\newblock {\em IEEE Transactions on Affective Computing}, 3(1):18--31, January
  2012.
\newblock Conference Name: IEEE Transactions on Affective Computing.

\bibitem{kwon_emotion_2021}
Jangho Kwon, Jihyeon Ha, Da-Hye Kim, Jun~Won Choi, and Laehyun Kim.
\newblock Emotion {Recognition} {Using} a {Glasses}-{Type} {Wearable} {Device}
  via {Multi}-{Channel} {Facial} {Responses}.
\newblock {\em IEEE Access}, 9:146392--146403, January 2021.
\newblock Publisher: IEEE.

\bibitem{lee_current_2021}
Seunggyu Lee, Hyewon Kim, Mi~Jin Park, and Hong~Jin Jeon.
\newblock Current {Advances} in {Wearable} {Devices} and {Their} {Sensors} in
  {Patients} {With} {Depression}.
\newblock {\em Frontiers in Psychiatry}, 12:672347, June 2021.

\bibitem{lejuez_modified_2003}
C.~W. Lejuez, Christopher~W. Kahler, and Richard~A. Brown.
\newblock A modified computer version of the {Paced} {Auditory} {Serial}
  {Addition} {Task} ({PASAT}) as a laboratory-based stressor.
\newblock {\em the Behavior Therapist}, 26(4):290--293, 2003.

\bibitem{liu_adolescent_1997}
Xianchen Liu, Lianqi Liu, Jie Yang, Fuxun Chai, Aizhen Wang, Liangming Sun,
  Guifan Zhao, and Dengdai Ma.
\newblock The {Adolescent} {Self}-{Rating} {Life} {Events} {Checklist} and its
  reliability and validity.
\newblock {\em Chinese Journal of Clinical Psychology}, 5(1):34--36, 1997.

\bibitem{miranda_calero_wemac_2024}
Jose~A. Miranda~Calero, Laura Gutiérrez-Martín, Esther Rituerto-González,
  Elena Romero-Perales, Jose~M. Lanza-Gutiérrez, Carmen Peláez-Moreno, and
  Celia López-Ongil.
\newblock {WEMAC}: {Women} and {Emotion} {Multi}-modal {Affective} {Computing}
  dataset.
\newblock {\em Scientific Data}, 11:1182, October 2024.

\bibitem{ramaswamy_multimodal_2024}
Manju Priya~Arthanarisamy Ramaswamy and Suja Palaniswamy.
\newblock Multimodal emotion recognition: {A} comprehensive review, trends, and
  challenges.
\newblock {\em WIREs Data Mining and Knowledge Discovery}, 14(6):e1563, 2024.
\newblock \_eprint: https://onlinelibrary.wiley.com/doi/pdf/10.1002/widm.1563.

\bibitem{reinhardt_salivary_2012}
Tatyana Reinhardt, Christian Schmahl, Stefan Wüst, and Martin Bohus.
\newblock Salivary cortisol, heart rate, electrodermal activity and subjective
  stress responses to the {Mannheim} {Multicomponent} {Stress} {Test} ({MMST}).
\newblock {\em Psychiatry Research}, 198(1):106--111, June 2012.

\bibitem{saganowski_emognition_2022}
Stanisław Saganowski, Joanna Komoszyńska, Maciej Behnke, Bartosz Perz,
  Dominika Kunc, Bartłomiej Klich, Łukasz~D. Kaczmarek, and Przemysław
  Kazienko.
\newblock Emognition dataset: emotion recognition with self-reports, facial
  expressions, and physiology using wearables.
\newblock {\em Scientific Data}, 9(1):158, April 2022.
\newblock Number: 1 Publisher: Nature Publishing Group.

\bibitem{shu_wearable_2020}
Lin Shu, Yang Yu, Wenzhuo Chen, Haoqiang Hua, Qin Li, Jianxiu Jin, and Xiangmin
  Xu.
\newblock Wearable {Emotion} {Recognition} {Using} {Heart} {Rate} {Data} from a
  {Smart} {Bracelet}.
\newblock {\em Sensors (Basel, Switzerland)}, 20(3):718, January 2020.

\bibitem{spielberger_state-trait_2017}
Charles~D. Spielberger, Fernando Gonzalez-Reigosa, Angel Martinez-Urrutia, Luiz
  F.~S. Natalicio, and Diana~S. Natalicio.
\newblock The {State}-{Trait} {Anxiety} {Inventory}.
\newblock {\em Revista Interamericana de Psicología/Interamerican Journal of
  Psychology}, 5(3 \& 4), July 2017.
\newblock Number: 3 \& 4.

\bibitem{wang_reliability_2014}
Wenzheng Wang, Qian Bian, Yan Zhao, Xu~Li, Wenwen Wang, Jiang Du, Guofang
  Zhang, Qing Zhou, and Min Zhao.
\newblock Reliability and validity of the {Chinese} version of the {Patient}
  {Health} {Questionnaire} ({PHQ}-9) in the general population.
\newblock {\em General Hospital Psychiatry}, 36(5):539--544, 2014.

\bibitem{wijasena_survey_2021}
Hamidan~Z. Wijasena, Ridi Ferdiana, and Sunu Wibirama.
\newblock A {Survey} of {Emotion} {Recognition} using {Physiological} {Signal}
  in {Wearable} {Devices}.
\newblock In {\em 2021 {International} {Conference} on {Artificial}
  {Intelligence} and {Mechatronics} {Systems} ({AIMS})}, pages 1--6, April
  2021.

\bibitem{zhang_development_2019}
Xintong Zhang, Meng-Cheng Wang, Lingnan He, Luo Jie, and Jiaxin Deng.
\newblock The development and psychometric evaluation of the {Chinese} {Big}
  {Five} {Personality} {Inventory}-15.
\newblock {\em PLoS ONE}, 14(8):e0221621, August 2019.

\bibitem{zhao_emotion_2021}
Sicheng Zhao, Guoli Jia, Jufeng Yang, Guiguang Ding, and Kurt Keutzer.
\newblock Emotion {Recognition} {From} {Multiple} {Modalities}: {Fundamentals}
  and methodologies.
\newblock {\em IEEE Signal Processing Magazine}, 38(6):59--73, November 2021.
\newblock Conference Name: IEEE Signal Processing Magazine.

\end{thebibliography}

\end{document}